\documentclass[
aps,
pra,
reprint,
groupedaddress,
]{revtex4-1}

\usepackage [utf8]{inputenc}

\usepackage{amssymb,amsmath}
\usepackage[
]{graphicx}
\usepackage{xcolor}

\usepackage[english]{babel}

\usepackage{siunitx}
\usepackage{braket}
\usepackage{pgfplots}

\usetikzlibrary{arrows,shapes,backgrounds,calc,positioning,intersections}

\makeatletter
\newsavebox\myboxA
\newsavebox\myboxB
\newlength\mylenA
\newcommand*\xoverline[2][0.75]{%
    \sbox{\myboxA}{$\m@th#2$}%
    \setbox\myboxB\null
    \ht\myboxB=\ht\myboxA%
    \dp\myboxB=\dp\myboxA%
    \wd\myboxB=#1\wd\myboxA
    \sbox\myboxB{$\m@th\overline{\copy\myboxB}$}
    \setlength\mylenA{\the\wd\myboxA}
    \addtolength\mylenA{-\the\wd\myboxB}%
    \ifdim\wd\myboxB<\wd\myboxA%
       \rlap{\hskip 0.5\mylenA\usebox\myboxB}{\usebox\myboxA}%
    \else
        \hskip -0.5\mylenA\rlap{\usebox\myboxA}{\hskip 0.5\mylenA\usebox\myboxB}%
    \fi}
\makeatother

\renewcommand{\ket}[1]{\left| #1 \right>} 
\renewcommand{\bra}[1]{\left< #1 \right|} 

\newcommand{\omc}{\omega_{\text{c}}}
\newcommand{\pr}{p_{\mathrm{rec}}}
\renewcommand{\vr}{v_{\mathrm{rec}}}
\newcommand{\Er}{E_{\mathrm{rec}}}

\newcommand{\dd}{\mathrm{d}}
\newcommand{\I}{\mathrm{i}}
\newcommand{\E}{\mathrm{e}}

\usepackage{layouts}
\usepackage{float}

\newcounter{Par}

\newcounter{Fig}

\definecolor{Blue}{cmyk}{1,0.9,0.,0.}
\definecolor{Red}{cmyk}{0.04,0.87,0.89,0}
\definecolor{Green}{cmyk}{0.75,0.,0.99,0}

\pgfplotsset{colormap={CM}{color=(white) color=(Blue!50!white) color=(Blue)  color=(Blue!75!black) color=(Blue!50!black)  color=(Blue!25!black) color=(black)}} 

\pgfplotsset{colormap={CM2}{color=(white) color=(Blue!25!white) color=(Blue!50!white) color=(Blue!75!white) color=(Blue)  color=(Blue!75!black) color=(Blue!50!black)  color=(Blue!25!black) color=(black)}} 

\pgfplotsset{colormap={CM3}{color=(white) color=(Red!50!white) color=(Red)  color=(Red!75!black) color=(Red!50!black)  color=(Red!25!black) color=(black)}} 

\pgfplotsset{colormap={CM4}{color=(white) color=(Red!50!white) color=(Red)   color=(Red!50!black) }} 

\pgfplotsset{colormap={BlueToRed}{color=(Blue)  color=(Red)}}

\graphicspath{
	{Figures_main/}
	{Figures_extended/}
	{Figures_supp/}
}

\usepackage{bibunits}
\usepackage{etoolbox}
\makeatletter
\newcommand*{\newbibstartnumber}[1]{%
  \apptocmd{\thebibliography}{%
    \global\c@NAT@ctr #1\relax
    \addtocounter{NAT@ctr}{-1}%
  }{}{}%
}
\makeatother

\begin{document}
\defaultbibliography{references}
\defaultbibliographystyle{naturemag}

\makeatletter
\def\fnum@figure{\figurename\nobreakspace\textbf{\thefigure}}
\makeatother

\renewcommand{\figurename}{\textbf{Fig.}}
\renewcommand\thefigure{\arabic{figure}}
\newcommand{\figref}[1]{Fig.\,\ref{#1}}

\title{Probing chiral edge dynamics and bulk topology of a synthetic  Hall system}

\author{Thomas Chalopin}
\thanks{These two authors contributed equally.}
\author{Tanish Satoor}
\thanks{These two authors contributed equally.}
\author{Alexandre Evrard}
\author{Vasiliy Makhalov}
\thanks{Present address: ICFO-The Institute of Photonic Sciences, 08860 Castelldefels (Barcelona), Spain}
\author{Jean Dalibard}
\author{Raphael Lopes}
\author{Sylvain Nascimbene}
\email{sylvain.nascimbene@lkb.ens.fr}

\affiliation{Laboratoire Kastler Brossel,  Coll\`ege de France, CNRS, ENS-PSL University, Sorbonne Universit\'e, 11 Place Marcelin Berthelot, 75005 Paris, France}
\date{\today}

\maketitle

\begin{bibunit}


\noindent\textbf{
Quantum Hall systems are characterized by the quantization of the Hall conductance -- a bulk property rooted in the topological structure of the underlying quantum states \cite{thouless_quantized_1982}.
In condensed matter devices, material imperfections hinder a direct connection to simple topological models \cite{laughlin_quantized_1981,halperin_quantized_1982}. 
Artificial systems, such as photonic platforms \cite{ozawa_topological_2019-1} or cold atomic gases \cite{goldman_topological_2016}, open novel possibilities by enabling  specific probes of topology \cite{jotzu_experimental_2014,aidelsburger_measuring_2015,hu_measurement_2015,mittal_measurement_2016,wu_realization_2016,flaschner_experimental_2016,ravets_polaron_2018,schine_electromagnetic_2019} or flexible manipulation e.g. using synthetic dimensions \cite{celi_synthetic_2014-1,mancini_observation_2015,stuhl_visualizing_2015,livi_synthetic_2016,kolkowitz_spinorbit-coupled_2017,an_direct_2017,lustig_photonic_2019,ozawa_topological_2019}.
However, the relevance of topological properties requires the notion of a bulk, which was missing in previous works using synthetic dimensions of limited sizes.
Here, we realize a quantum Hall system using ultracold dysprosium atoms, in a two-dimensional geometry formed by one spatial dimension and one synthetic dimension encoded in the atomic spin $J=8$.
We demonstrate that the large number of magnetic sublevels leads to distinct bulk and edge behaviors. 
Furthermore, we measure the Hall drift and reconstruct the local Chern marker, an observable that has remained, so far, experimentally inaccessible \cite{bianco_mapping_2011}.
In the center of the synthetic dimension -- a bulk of 11 states out of 17 -- the Chern marker reaches 98(5)\% of the quantized value expected for a topological system.
Our findings pave the way towards the realization of topological many-body phases.
}

In two-dimensional electron gases, the quantization of the Hall conductance results from the non-trivial topological structuring of the quantum states of an electron band.
For an infinite system, this topological character is described by the Chern number $\mathcal{C}$, a global invariant taking a non-zero integer value that is robust to relatively weak disorder  \cite{thouless_quantized_1982}.
In a real finite-size system, the non-trivial topology further leads to in-gap excitations delocalized over the edges, characterized by unidirectional motion exempt from backscattering \cite{halperin_quantized_1982}.
Such protected edge modes, together with their generalization to topological insulators, topological superconductors or fractional quantum Hall states \cite{stormer_fractional_1999,hasan_colloquium_2010},  lie at the heart of possible applications in spintronics \cite{pesin_spintronics_2012} or quantum computing \cite{kitaev_fault-tolerant_2003}.

In electronic quantum Hall systems, the topology manifests itself via the spectacular robustness of the Hall conductance quantization to finite-size or disorder effects \cite{klitzing_new_1980}.
Nonetheless, such perturbations typically lead to conducting stripes surrounding insulating domains of localized electrons, making the comparison with simple defect-free models challenging.
In topological insulators or fractional quantum Hall systems, topological properties are more fragile, and can only be revealed in very clean samples \cite{stormer_fractional_1999,hasan_colloquium_2010}.
Recent experiments with topological quantum systems in photonic or atomic platforms \cite{lu_topological_2014,goldman_topological_2016} have created new possibilities, from the realization  of emblematic models of topological matter \cite{aidelsburger_realization_2013,miyake_realizing_2013,jotzu_experimental_2014} to the application of well-controlled  edge and disorder potentials.
In such systems, internal degrees of freedom can be used to simulate a synthetic dimension of finite size with sharp-edge effects \cite{celi_synthetic_2014-1,mancini_observation_2015,stuhl_visualizing_2015,livi_synthetic_2016,kolkowitz_spinorbit-coupled_2017,an_direct_2017,lustig_photonic_2019,ozawa_topological_2019}.
Encoding a synthetic dimension in the time domain can also give access to higher-dimensional physics \cite{lohse_exploring_2018,zilberberg_photonic_2018}.

In this work, we engineer a topological system with ultracold bosonic $^{162}$Dy atoms based on coherent light-induced couplings between the atom's motion and the electronic spin $J=8$, with relevant dynamics along two dimensions -- one spatial dimension and a synthetic dimension encoded in the discrete set of $2J+1=17$ magnetic sublevels.
These couplings give rise to an artificial magnetic field, such that our system realizes an analog of a quantum Hall ribbon.
In the lowest band, we characterize the dispersionless bulk modes, where motion is inhibited due to a flattened energy band, and edge states, where the particles are free to move in one direction only.
We also study elementary excitations to higher bands, which take the form of cyclotron and skipping orbits.
We furthermore measure the Hall drift induced by an external force, and infer the local Hall response of the band via the local Chern marker, which quantifies topological order in real space \cite{bianco_mapping_2011}. 
Our experiments show that the synthetic dimension is large enough to allow for a meaningful bulk with robust topological properties.
Numerical simulations of interacting bosons moreover show that our system can host quantum many-body systems with non-trivial topology, such as mean-field Abrikosov vortex lattices or fractional quantum Hall states.

\begin{figure*}[!ht]
\includegraphics[
draft=false,scale=1,
trim={3mm 2mm 0 0.cm},
]{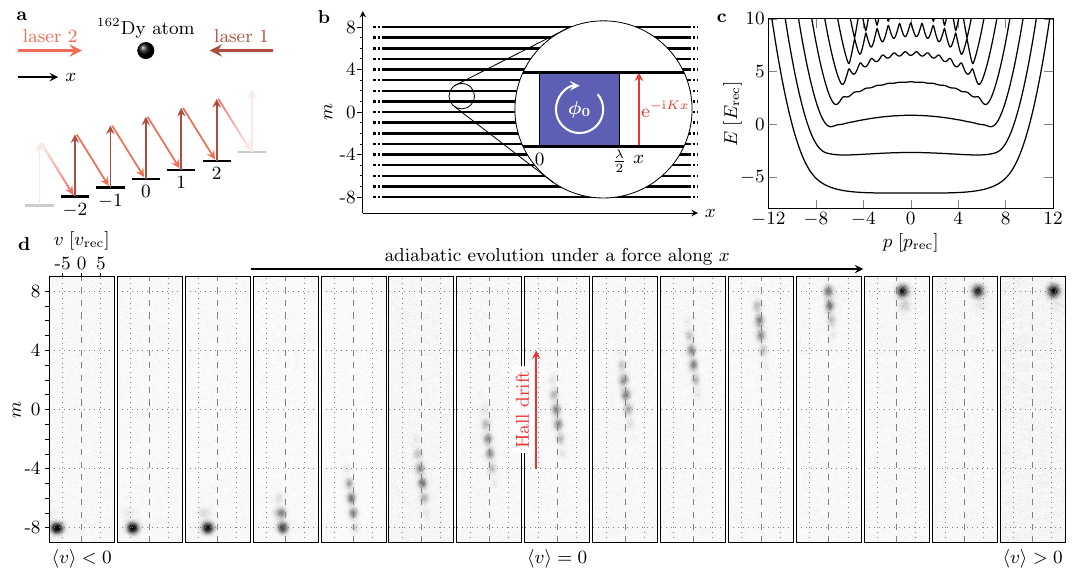}
\caption{
\textbf{Synthetic Hall system.}
\textbf{a.} Laser configuration used to couple the magnetic sublevels $m$ of a $^{162}$Dy atom (with $-J\leq m\leq J$ and $J=8$, only a few $m$ values represented).
\textbf{b.} Interpreting the spin projection as a  synthetic dimension, the system is mapped to a two-dimensional ribbon of finite width.
The photon recoil $p_{\mathrm{rec}}=\hbar K$ imparted upon a spin transition leads to complex-valued hopping amplitudes along $m$, equivalent to the Aharonov-Bohm phase of a charged particle evolving in a magnetic field.
The blue area represents a magnetic unit cell pierced by one  flux quantum $\phi_0$.
\textbf{c.} Dispersion relation describing the quantum level structure for $\hbar \Omega = \Er$, with flattened energy bands reminiscent of Landau levels.
\textbf{d.} The lowest energy band is explored by applying an external force.
We probe the velocity and magnetic projection distributions by imaging the atomic gas after an expansion under a magnetic field gradient.
We find three types of behavior: free motion with negative (positive) velocity on the bottom edge $m=-J$ (top edge $m=J$) and zero average velocity in the bulk.
Each panel corresponds to a single-shot image.
\label{fig_scheme}}
\end{figure*}

The atom dynamics is induced by two-photon optical transitions involving counter-propagating laser beams  along $x$ (see \figref{fig_scheme}a), and coupling successive magnetic sublevels $m$ \cite{lin_spinorbit-coupled_2011,cui_synthetic_2013}.
Here, the integer $m$ ($-J\leq m\leq J$) quantifies the spin projection along the direction $z$ of an external magnetic field.
The spin coupling amplitudes then inherit the complex phase $Kx$ of the interference between both lasers, where $K=4\pi/\lambda$ and $\lambda=\SI{626.1}{nm}$ is the light wavelength (see \figref{fig_scheme}b).
Given the Clebsch-Gordan algebra of atom-light interactions for the dominant optical transition, the atom dynamics is described by the Hamiltonian
\begin{equation}
\hat H=\frac{1}{2}M\hat v^2-\frac{\hbar\Omega}{2}\left(\E^{-\I K\hat x}\hat J_+ + \E^{\I K\hat x}\hat J_-\right) + V(\hat J_z)\label{eq_H0}
\end{equation}
where $M$ is the atom mass, $\hat v$ is its velocity, $\hat J_z$ and $\hat J_\pm$ are spin projection and ladder operators.
The coupling  $\Omega$ is proportional to both laser electric fields, and the potential $V(\hat J_z)=-\hbar\Omega\hat J_z^2/(2J+3)$ stems from rank-2 tensor light shifts (see Methods and Supplementary information).

\begin{figure*}[!ht]
\includegraphics[
draft=false,scale=1,
trim={1mm 4mm 0 1mm},
]{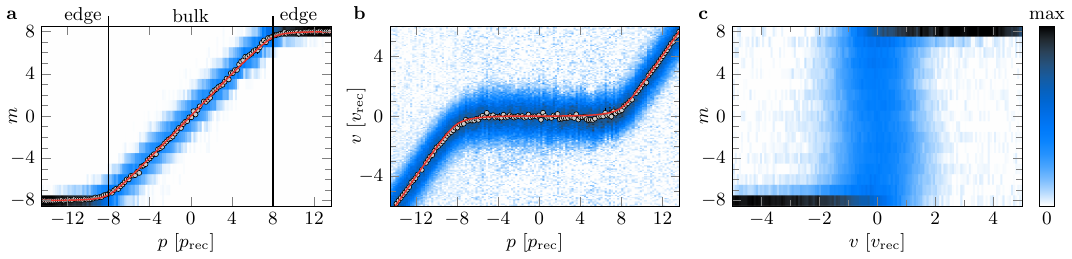}
\caption{
\textbf{Ground band characterization.}
\textbf{a.} Spin projection distribution $\Pi_m$ as a function of the momentum $p$,  with the mean spin projection $\langle \hat J_z\rangle$ (gray dots) and the theoretical prediction $(p-M\partial_pE_0)/\pr$ (red line).
\textbf{b.} Velocity distribution, together with the mean velocity $\langle \hat v \rangle$ (gray dots) and the expected value, given by  the derivative of the band dispersion $\partial_p E_0$ (red line).
\textbf{c.} Local density of states obtained by integrating the measured distributions in $(v,m)$ space over all momenta.
All error bars are the 1-$\sigma$ standard deviation of typically 5 measurement repetitions.
\label{fig_ground_state}}
\vspace{-0.5em}
\end{figure*}

A light-induced spin transition $m\rightarrow m+1$ is accompanied by a momentum kick $-\pr\equiv-\hbar K$ along $x$, such that  the canonical momentum $\hat p=M \hat v+\pr \hat J_z$ is a conserved quantity.
After a unitary transformation defined by the operator $\hat U=\exp(\I K\hat x\hat J_z)$, the Hamiltonian (\ref{eq_H0}) can be rewritten, for a given momentum $p$, as
\begin{equation}
	\hat H_p=\frac{(p- \pr  \hat J_z)^2}{2M}-\hbar\Omega \hat J_x+V(\hat J_z)\, .
	\label{eq_H}
\end{equation}
We can make an analogy between this Hamiltonian and the \textit{ideal} Landau one, given by
\begin{equation}
	\hat H_{\mathrm{Landau}}=\frac{( \hat p_x-e B  \hat y)^2}{2M}+\frac{ \hat p_y^2}{2 M},
	\label{eq_H_Landau}
\end{equation}
which describes the dynamics of an electron evolving in 2D under a  perpendicular magnetic field $B$.
The analogy between both systems can be made upon the identifications $ \hat J_z\leftrightarrow  \hat y$ and $\pr \leftrightarrow e B$.
The term $ \hat J_x$ in equation (\ref{eq_H}) plays the role of the kinetic energy along the synthetic dimension, since it couples neighboring $m$ levels with real positive coefficients, similarly to the discrete form of the Laplacian operator $\propto \hat p_y^2$ in equation (\ref{eq_H_Landau}) (see Supplementary Information).
The range of magnetic projections being limited, our system maps onto a Hall system in a ribbon geometry bounded by the edge states $m=\pm J$.
The relevance of the analogy  is confirmed by the structure of energy bands $E_n(p)$ expected for the Hamiltonian (\ref{eq_H}) describing our system, shown in \figref{fig_scheme}c.
The energy dispersion of the first few bands is strongly reduced  for $|p|\lesssim J \pr$, reminiscent of dispersionless Landau levels.
A parabolic dispersion is recovered for $|p|\gtrsim J\pr$, similar to the ballistic edge modes of a quantum Hall ribbon \cite{halperin_quantized_1982}.
The flatness of the lowest energy band, for $\hbar\Omega \approx \Er$, results from the compensation of two dispersive effects, namely the   variation of $\hat J_x$ matrix elements and the extra term, $V(\hat J_z)$ (see Supplementary information).

We first characterize  the ground band using quantum states of arbitrary values of momentum $p$.
We begin with an atomic gas spin-polarized in $m=-J$, and with a negative mean velocity $\langle \hat{v}\rangle=-6.5(1)\vr$ (with $\vr\equiv\pr/M)$, such that it corresponds to the ground state of (\ref{eq_H}) with $p=-14.5(1)\pr$.
The gas temperature $T=\SI{0.55(6)}{\micro K}$ is such that the thermal velocity broadening is smaller than the recoil velocity $\vr$.
We then slowly increase the light intensity up to a coupling $\hbar\Omega=1.02(6)\Er$, where $\Er\equiv\pr^2/(2M)$ is the natural energy scale in our system.
Subsequently, we apply a weak force $F_x$ along $x$, such that the state adiabatically evolves in the ground energy band with $\dot p=F_x$, until the desired momentum is reached (see Methods).
We measure the distribution of velocity $v$ and spin projection $m$ by imaging the atomic gas after a free flight in the presence of a magnetic field gradient.

The main features of Landau level physics are visible in the raw images shown in \figref{fig_scheme}d.
Depending on the momentum $p$, the system exhibits three types of behaviors.
(i) When spin-polarized in $m=-J$, the atoms move with a negative mean velocity $\langle \hat v\rangle$, consistent with a left-moving edge mode.
(ii) When the velocity approaches zero under the action of the force $F_x$, the system experiences a series of resonant transitions to higher $m$ sublevels -- in other words a transverse Hall drift along the synthetic dimension.
In this regime the atom's motion is inhibited along $x$, as expected for a quasi non-dispersive band.
(iii) Once the edge $m=J$ is reached, the velocity $\langle \hat v\rangle$ rises again, corresponding to a right-moving edge mode.
Overall, while exploring  the entire ground band under the action of a force along $x$, the atoms are pumped from one edge to the other along the synthetic dimension.

To distinguish between bulk and edge modes, we plot in \figref{fig_ground_state}a the spin projection probabilities $\Pi_m$ as a function of momentum $p$.
We find that the edge probabilities $\Pi_{m=\pm J}$ exceed 1/2 for $|p|>8.0(1)\pr$, defining the edge mode sectors -- with the bulk modes in between.
We study the  system dynamics via its velocity distribution and mean velocity $\langle\hat  v\rangle$, shown in \figref{fig_ground_state}b.
We observe that the velocity of bulk modes remains close to zero, which shows via the Hellmann-Feynman relation $\langle\hat  v\rangle=\partial E_0/\partial p$ that the ground band is almost flat.

 The measured residual mean velocities allow us to infer a dispersion $\Delta E_0^{\mathrm{pk-pk}}=1.2(5)\,\Er$ in the bulk mode region -- nearly $2\%$  of the free-particle dispersion expected over the same range of momenta.
On the contrary, edge modes are characterized by a velocity $\langle\hat  v\rangle\simeq (p-p_0)/M$, corresponding to ballistic motion -- albeit with the restriction $\langle \hat v\rangle<0$ for edge modes close to $m=-J$, and $\langle\hat v\rangle>0$ at the opposite edge.
We also characterize correlations between velocity $v$ and spin projection $m$ over the full band, via the local density of states (LDOS) in $(v,m)$ space, integrated over $p$.
We stress here that the LDOS only involves gauge-independent quantities, and could thus be generalized to more complex geometries lacking translational invariance.
As  shown in \figref{fig_ground_state}c, it also reveals characteristic quantum Hall behavior, namely inhibited dynamics in the bulk and chiral motion on the edges.

\begin{figure}[!t]
\includegraphics[
draft=false,scale=1,
trim={2mm 3mm 0 0.cm},
]{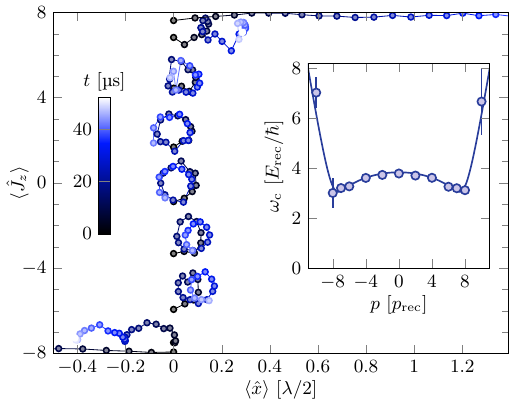}
\caption{
\textbf{Cyclotron and skipping orbits}.
Trajectories in $(x, m)$ space following a velocity kick $v_{\mathrm{kick}}\simeq\vr$, starting at $\langle \hat x\rangle=0$, and for different initial momentum states (blue dots).
The color encodes the time evolution.
Inset: Frequencies extracted from the velocity dynamics (blue dots), and compared with the expected cyclotron gap for $\hbar\Omega=\Er$ (blue line).
The error bars are the 1-$\sigma$ statistical uncertainty calculated from a bootstrap sampling analysis performed on more than a hundred pictures.
\label{fig_orbits}}
\vspace{-0.5em}
\end{figure}

The ideal Landau level structure of a charged particle evolving in two dimensions in a transverse magnetic field is characterized by a harmonic energy spacing $\hbar\omega_{\text c}$, set by the cyclotron frequency $\omega_{\text c}=eB/M$.
We test this behavior in our system by studying elementary excitations above the ground band, via the trajectories of the center of mass following a velocity kick $v_{\mathrm{kick}}\simeq\vr$.
To access the real-space position of the atoms, we numerically integrate their center-of-mass velocity evolution (see Methods).
As shown in \figref{fig_orbits} (blue dots), we measure almost-closed trajectories in the bulk, consistent with the periodic cyclotron orbits expected for an infinite Hall system.
We checked that this behavior remains valid for larger excitation strengths, until one couples to highly dispersive excited bands (for velocity kicks  $v_{\mathrm{kick}}\gtrsim2\vr$, see Methods).
Close to the edges, the atoms experience an additional drift and their trajectories are similar to classical skipping orbits bouncing on a hard wall.
In particular, the drift orientation only depends on the considered edge, irrespective of the kick direction.
We report in the inset of \figref{fig_orbits} the frequencies of velocity oscillations, which agree well with the expected cyclotron gap to the first excited band.
We find that the gap is almost uniform within the bulk mode sector, with a residual variation in the range $\omega_c=3.0(1)-3.8(1)\Er/\hbar$.

\begin{figure}[t!]
\includegraphics[
draft=false,scale=1,
trim={4mm 4mm 0 0.cm},
]{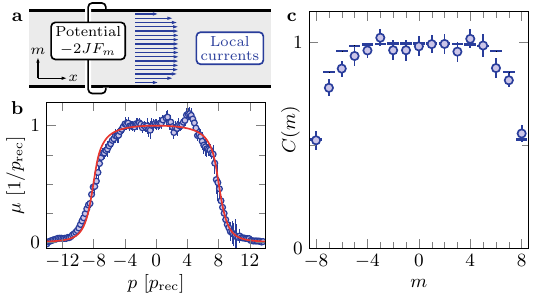}
\caption{
\textbf{Hall response}.
\textbf{a.} We determine the Hall response from the measurement of local currents in the real dimension, that result from the application, in synthetic space, of a potential difference $-2JF_m$ between the edges.
\textbf{b.} Hall mobility $\mu(p)$ measured for states of momenta $p$, via their increase of velocity upon a small force $F_m$ along $m$.
\textbf{c.} Local Chern marker as a function of $m$, corresponding to the integrated mobility $\mu(p)$ weighted by the projection probability $\Pi_m(p)$.
The error bars are the 1-$\sigma$ statistical uncertainty calculated from a bootstrap sampling analysis over typically 100 pictures (\textbf{b}) and 1000 pictures (\textbf{c}).
\label{fig_hall}}
\vspace{-0.5em}
\end{figure}

We now investigate the key feature of Landau levels, namely their quantized Hall response, which is intrinsically related to their topological nature.
In a ribbon geometry, the Hall response of a particle corresponds to the transverse velocity acquired upon applying a potential difference across the edges (see \figref{fig_hall}a).
In our system, such a potential corresponds to a Zeeman term $-F_m  \hat J_z$ added to the Hamiltonian (\ref{eq_H}), which can now be recast as
\[
	\hat H_p-F_m  \hat J_z= \hat H_{p+M v'}-v' p,\quad\text{with}\quad v'=F_m/\pr,
\]
such that the force acts as a momentum shift $M v'$ in the reference frame with velocity $v'$.
In the weak force limit, the perturbed state remains in the ground band, and its  mean velocity reads
\[
	\langle\hat v\rangle=\langle\hat v\rangle_{F_m=0}- \mu F_m,\quad \text{where}\quad\mu=\frac{1}{\pr}\frac{\partial}{\partial p}(p-M\langle\hat v\rangle)
\]
is the Hall mobility.
This expression shows that the Hall response to a weak force can be related to the variation of the mean velocity within the ground band, that we show in \figref{fig_ground_state}b.
In practice, the velocity derivative  at momentum $p$ is evaluated using momentum states in the domain $(p-\pr,p+\pr)$, corresponding to an evaluation of the Hall drift under a force $-2\,\Er/\ell<F_m<2\,\Er/\ell$, where $\ell =1$ is the unit length along the synthetic dimension.
We present in \figref{fig_hall}b the Hall mobility $\mu(p)$ deduced from this procedure.
For bulk modes, it remains close to the value $\mu=1/\pr$, which corresponds to the classical mobility $\mu=1/(eB)$ in the equivalent Hall system.
The mobility decreases in the edge mode sector, as expected for topologically protected boundary states whose ballistic motion is  undisturbed by the magnetic field.

We use the measured drift of individual quantum states to infer the overall Hall response of the ground band.
As for any spatially limited sample, our system does not exhibit a gap in the energy spectrum due to the edge mode dispersion.
In particular, high-energy edge modes of the ground band are expected to resonantly hybridize with excited bands upon disorder, such that defining the Hall response of the entire ground band is not physically meaningful.
We thus only consider the energy branch $E<E^*$, where $E^*$ lies in the middle of the first gap at zero momentum (see Methods).
We characterize the (inhomogeneous) Hall response of this branch via  the local Chern marker 
\[
	C(m)\equiv 2\pi\,\text{Im}\bra{m}[ \hat P  \hat x \hat P, \hat P  \hat J_z \hat P]\ket{m}=\int_{E(p)<E^*}\!\!\!\!\!\!\!\!\!\!\!\!\!\!\!\text{d}p\,\Pi_m(p)\mu(p),
\]
where $\hat P$ projects on the chosen branch \cite{kitaev_anyons_2006,bianco_mapping_2011}.
This local geometrical marker quantifies the adiabatic transverse response in position space, and matches the integer Chern number $\mathcal{C}$ in the bulk of a large, defect-free system.
Here, it is given by the integrated mobility $\mu(p)$, weighted by the spin projection probability $\Pi_m(p)$ (see Methods).
As shown in \figref{fig_hall}c, we identify a plateau in the range $-5\leq m\leq 5$.
There, the average value of the Chern marker, $\xoverline{C}=0.98(5)$, is consistent with the integer value $\mathcal{C}=1$  -- the Chern number of an infinite Landau level.
This measurement shows that our system is large enough to reproduce  a  topological Hall response in its bulk.
For positions $|m|\geq 6$, we measure a decrease of the Chern marker, that we attribute to non-negligible correlations with the edges.

\begin{figure}[t!]
\includegraphics[
draft=false,scale=1,
trim={2mm 1mm 0 0.cm},
]{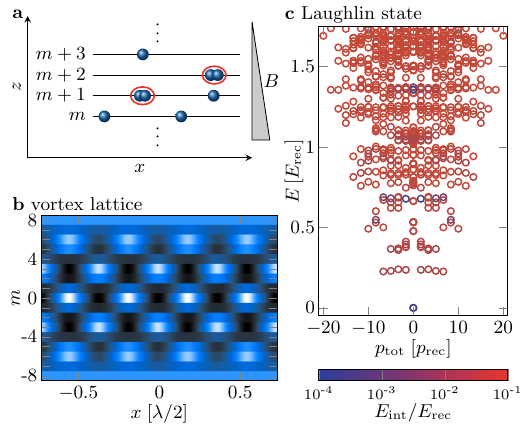}
\caption{
\textbf{Simulations of topological many-body systems.}
\textbf{a.} Proposed scheme to engineer contact interactions along both directions $x$ and $m$.
A magnetic field gradient spatially separates the different $m$ states along $z$, such that collisions (represented by the red ellipses) only occur for atoms in the same magnetic sublevel $m$.
\textbf{b.} Density distribution of a Bose-Einstein condensate, with a chemical potential at $\simeq 2\Er$ above the single-particle ground state energy.
\textbf{c.} Many-body spectrum of a system of $N=5$ interacting atoms, where the color encodes the interaction energy.
We use periodic boundary conditions along $x$, with a circumference $L=0.6(\lambda/2)$ allowing for $N_{\text{orb}}=9$ orbitals at low energy, compatible with the Laughlin state.
The residual energy dispersion between these orbitals is minimized by using a coupling strength $\Omega=0.5\,\Er/\hbar$.
\label{fig_vortex}}
\vspace{-0.5em}
\end{figure}

Such a topological bulk is a prerequisite for the realization of emblematic phases of two-dimensional quantum Hall systems, as we now confirm via numerical simulations of interacting quantum many-body systems.
In our system, collisions between atoms \emph{a priori} occur when they are located at the same position $x$, irrespective of their spin projections $m$, $m'$, leading to highly anisotropic interactions.
While this feature leads to an interesting phenomenology \cite{barbarino_magnetic_2015}, we propose to control the interaction range by spatially separating the different $m$ states using a magnetic field gradient,  suppressing both contact and dipole-dipole interactions for $m \neq m'$, as illustrated in \figref{fig_vortex}a (see Methods and Supplementary information).
The system then becomes truly two-dimensional, and closely related to the seminal work of \cite{lin_spinorbit-coupled_2011}, albeit with a discrete spatial dimension with sharp walls.
We discuss below the many-body phases expected for bosonic atoms with such short-range interactions, assuming for simplicity repulsive interactions of equal strength for each projection $m$.

We first consider the case of a large filling fraction $\nu\equiv N_{\text{at}}/N_{\phi}\gg 1$, where $N_{\phi}$ is the number of magnetic flux quanta in the area occupied by $N_{\text{at}}$ atoms -- as realized in previous experiments on rapidly rotating gases \cite{schweikhard_rapidly_2004,bretin_fast_2004}.
In this regime and at low temperature, the system forms  a Bose-Einstein condensate that spontaneously breaks translational symmetry, leading to a triangular Abrikosov lattice of quantum vortices (see \figref{fig_vortex}b).
Due to the hard-wall boundary along $m$, one expects  phase transitions between  vortex lattice configurations when tuning the coupling strength $\Omega$ and the chemical potential, similar to the phenomenology of type-II superconductors in confined geometries \cite{abrikosov_magnetic_1957} (see Methods).

For lower filling fractions $\nu\sim1$, one expects strongly-correlated ground states analogous to fractional quantum Hall states \cite{kane_fractional_2002}.
We present in \figref{fig_vortex}c a numerical calculation of the many-body spectrum for $N_{\text{at}}=5$ atoms with periodic boundary conditions along $x$, corresponding to a cylinder geometry.
We choose the circumference such that the number of orbitals $N_{\text{orb}}=9$ in the bulk region of the lowest band matches the number $N_{\text{orb}}=2N_{\text{at}}-1$ required to construct the Laughlin wavefunction.
For contact interactions parametrized by a Haldane pseudopotential of amplitude $U=\Er$, we numerically find  a ground state separated by an energy gap $E_{\text{gap}}\simeq0.23\,\Er=k_{\text{B}}\times\SI{140}{nK}$ from the rest of the excitations.
It also exhibits a very small interaction energy $E_{\text{int}}$, indicating anti-bunching between atoms, which is a hallmark of the Laughlin state.

The realization of a quantum-Hall system based on a large synthetic dimension, as discussed here, is a promising setting for  future realizations of topological quantum matter.
An important asset of our setup is the large cyclotron energy, measured in the range $\hbar\omega_{\text c}\simeq k_{\text{B}}\times 1.8-2.3\,\si{\micro K}$, much larger than the typical temperatures of quantum degenerate gases, thus enabling the realization of strongly-correlated states  at realistic temperatures.
The techniques developed here could give access to complex correlation effects, such as flux attachment via cyclotron orbits \cite{goldman_detection_1994} or charge fractionalization via adiabatic pumping \cite{zeng_charge_2015} or center-of-mass Hall response \cite{taddia_topological_2017}.
Our protocol could also be extended to fermionic isotopes of dysprosium, with a synthetic dimension given by the hyperfine spin of the lowest energy state, $F=21/2$ for $^{161}$Dy, leading to an even larger bulk.
At low temperature and unit filling of the ground band the Fermi sea would exhibit an almost quantized Hall response akin to the integer quantum Hall effect.

\putbib
\end{bibunit}
 
\smallskip
\noindent\textbf{Acknowledgements}\\
\noindent 
We thank J. Beugnon, N. Cooper, P. Delpace, N. Goldman, L. Mazza, and H. Price  for stimulating discussions.
We acknowledge funding by the EU under the ERC projects `UQUAM' and `TOPODY', and PSL research university under the project `MAFAG'.

\smallskip
\noindent\textbf{Author contributions}\\
\noindent 
All authors contributed to the setup of the experiment, the data acquisition, its analysis and the writing of the manuscript.

\smallskip
\noindent\textbf{Competing interests}\\
\noindent 
The authors declare no competing interests.

\smallskip
\noindent\textbf{Author Information}\\
\noindent 
Correspondence and requests for materials should be addressed to S.N. (sylvain.nascimbene@lkb.ens.fr).

\cleardoublepage

\section*{Methods}

\begin{bibunit}[naturemag]

\smallskip
\noindent\textbf{Details on the experimental protocol}\\
	\noindent Our experiments begin by preparing an ultracold gas of $8(2)\times 10^4$ $^{162}$Dy atoms at a temperature $T=\SI{0.55(6)}{\micro\kelvin}$, and held in an almost symmetrical optical dipole trap with frequency $\bar{\omega} = 2\pi\times 150$~Hz, leading to a peak density of $n_0 \approx 10^{13}$~cm$^{-3}$.
	The atoms are placed in a magnetic field $B=\SI{172(2)}{mG}$ along the $z$-axis, corresponding to a Zeeman splitting of frequency $\omega_{\text{Z}}=2\pi\times\SI{298(3)}{kHz}$, with the electronic spin polarized in the absolute ground state $m=-J$.
	We then turn off the trap, and turn on the two laser beams shown in Fig.\,\ref{fig_scheme}, which differ in frequency by $\omega_{12}=\omega_1-\omega_2$.
	When $\omega_{12}$ is close to the Zeeman splitting $\omega_{\text{Z}}$,  a  spin transition $m\rightarrow m+1$ occurs via the absorption of one photon from beam 1 and the stimulated emission of one photon in beam 2.
	In such processes and in the absence of additional external forces, the canonical momentum $\hat p=M\hat v+\hbar K \hat J_z$ is conserved.

	The laser beam frequencies are set close to the optical transition at $\SI{626.1}{nm}$, which couples the electronic ground state $J=8$ to an excited level $J'=9$.
	The beams are detuned by $\Delta=2\pi\times\SI{22}{\giga\hertz}$ with respect to resonance and are linearly polarized along orthogonal directions, each being at $\SI{45}{\degree}$ with respect to the $z$-axis.
	Then, the algebra of Clebsch-Gordan coefficients of $J\rightarrow J'=J+1$ transitions leads to the Hamiltonian \eqref{eq_H} at resonance ($\omega_{12}=\omega_{\text{Z}}$), with
	\[
	\Omega=\frac{2J+3}{4(J+1)(2J+1)}V_0,\quad V_0=\frac{3\pi c^2\Gamma}{2\omega_0^3}\frac{\sqrt{I_1I_2}}{\Delta}
	\] where $I_{1,2}$ are the laser intensities on the atoms, $\Gamma\simeq2\pi\times\SI{135}{\kilo\hertz}$ is the transition linewidth, and $\omega_0$ is its resonant frequency.
	The value of the coupling $\Omega$ is calibrated using an independent method and remains constant over the experimental sequence since the waists of both laser beams are much larger than the region of atomic motion.
	The Larmor frequency $\omega_{\text{Z}}$ is calibrated from the resonance of the Raman transition between $m=-8$ and $m=-7$.

\begin{figure}[t!]
\includegraphics[
draft=false,scale=1.,
trim={1mm 2mm 0 0.cm},
]{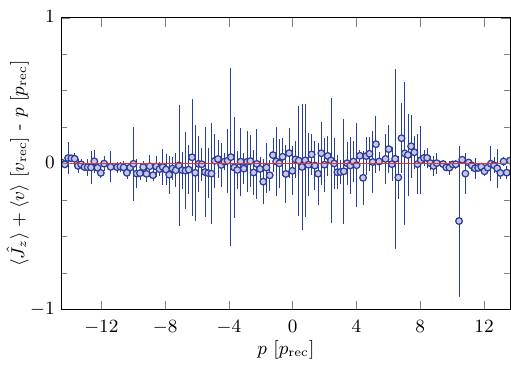}
\caption{
\textbf{Conservation of canonical momentum.}
Difference between the measured canonical momentum $\pr\langle \hat J_z\rangle+M\langle\hat v\rangle$ and the targeted value $p$ defined by the state preparation protocol.
All error bars are the 1-$\sigma$ standard deviation of typically 5 measurement repetitions.
\label{conservation_p}
}
\end{figure}

	The non-resonant case ($\omega_{12}\neq\omega_{\text{Z}}$)  can be reduced to the resonant case in a reference frame moving at a velocity $v_{\text{frame}}=(\omega_{\text{Z}}-\omega_{12})/K$.
	Note that the required change of frame means that fluctuations of $\omega_{\text{Z}}$ contribute to the uncertainties of the measured velocities.
	We first slowly increase the intensity up to a coupling $\hbar\Omega=1.02(6)\Er$, where $\Er\equiv\pr^2/(2M)$, and then apply an external force $F_x$ on the system via the inertial force resulting from a time-dependent frequency difference, with $F_x=(M/K)\partial_t\omega_{12}$.
	The preparation of a state in the lowest band with a given momentum $p$ is performed by adiabatically ramping the frequency difference to a final value
	\begin{equation}
		\omega_{12}=\omega_{\text{Z}}+2\left(\frac{p}{\pr}+J\right)\frac{\Er}{\hbar}.
		\label{def_p}
	\end{equation}
	We use the relation (\ref{def_p}) to define, from the final frequency difference $\omega_{12}$, the quasi-momentum $p$ parametrizing the experimental data.
	We use a constant ramp rate $\partial_t\omega_{12}\simeq 0.22\,\omega_{\text{c,min}}^2$, where $\omega_{\text{c,min}}\simeq3.06\,\Er/\hbar$ is the minimum cyclotron frequency separating the two lowest energy bands for $\Omega=\Er/\hbar$.
	Depending on the target $p$ state, the preparation takes between $\SI{150}{\micro\second}$ and $\SI{550}{\micro\second}$.
	Shot-to-shot fluctuations of the ambient magnetic field induce fluctuations of the Zeeman frequency splitting, hence an error in the value of the prepared momentum $p$.
	As shown in  \figref{conservation_p}, the measured error in momentum remains small compared to the recoil momentum $\pr$, and its rms deviation $\Delta p\simeq 0.06\,\pr$ is compatible with magnetic field fluctuations $\sigma_B=\SI{0.7}{mG}$ measured independently.
	
	We numerically checked the adiabaticity of the state preparation protocol.
	While preparing $m=+J$, which requires crossing all momentum states, the squared overlap with the ground band remains greater than 0.96 and the deviation of the mean spin projection $\langle \hat J_z\rangle$ from the corresponding ground state value is always less than 0.08.
	The largest deviations occur near $m\simeq \pm 7$, where the energy gap to the first excited band is the smallest.
	This behavior is consistent with our measurements, showing that the adiabatic transfer to $m=J$ after exploring the entire band is above 97\%.
	
	\begin{figure*}[t!]
\includegraphics[
draft=false,scale=1.,
trim={1mm 2mm 0 0.cm},
]{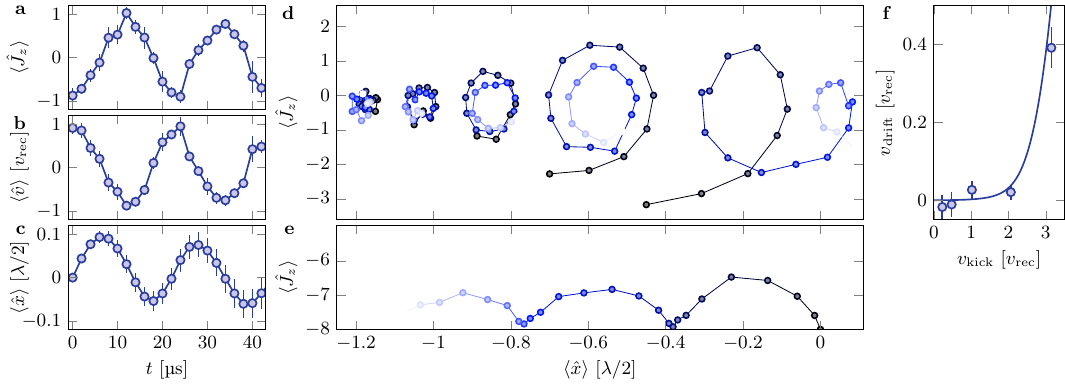}
\caption{
\textbf{Cyclotron orbits measurements.}
\textbf{a, b, c.} Magnetization, velocity and position response as a function of time after application of a velocity kick $v_\text{kick} \approx v_\text{rec}$.
\textbf{d.} Bulk excitations corresponding to different velocity kicks, $v= 0.23,\, 0.48,\, 1.02,\, 2.06,\, \text{and}\; 3.15\; v_\text{rec}$, from left to right.
The orbits are off-centred in real space for visual clarity.
\textbf{e.} Skipping orbit for the momentum state $p=-J\,\pr$ following a sudden jump of the coupling strength $\Omega$.
\textbf{f.} Velocity drift of the orbits as a function of the amplitude kick.
The solid line corresponds to the expected drift due to the non-harmonic spectrum of the energy bands.
All error bars are the 1-$\sigma$ standard deviation of typically 5 measurement repetitions.
\label{fig_orbits_radii}}
\end{figure*} 

	At the end of the experiment, we probe the velocity and spin projection distributions.
	For this, we abruptly switch off the Raman lasers, and subsequently ramp up an inhomogeneous magnetic field that splits the different magnetic sublevels along $z$.
	After a 4 ms expansion, we take a resonant absorption picture.
	The measured atom density is split along $z$ according to the magnetic projection $m$, and the density along $x$ corresponds to the distribution of velocity $v$ ($\omega t_\text{tof}\sim 4$).
	Our imaging setup is such that the 17 magnetic sublevels have different cross-sections.
	We calibrate the relative cross-sections such that the calculated atom number remains constant for all momentum states, irrespective of their spin composition.


\smallskip
\noindent\textbf{Cyclotron orbits} \\
	\noindent In order to probe the excitations of the system we perform a velocity quench, which couples the lowest Landau level to the next higher energy band.
	The system then responds periodically with a frequency set by the energy difference between the two bands, which for the case of an ideal Hall system would correspond to the cyclotron frequency $\omega_\text{c}$.
	Experimentally, we perform the velocity kick by quenching the detuning $\omega_{12}$, which in practice settles to a steady value after $\SI{4}{\micro s}$.
	We show in \figref{fig_orbits_radii}a,b an example of coherent oscillations of both magnetization and velocity.
	We compute the response of the system in real space, $x$, via a numerical integration of the velocity evolution as shown in \figref{fig_orbits_radii}c.
	The uncertainty on the Larmor frequency leads to a systematic error on the velocity on the order of $0.1\,\vr$, consistent with the small drift of some cyclotron orbits in the bulk.

	The response of the system is probed after a velocity kick $v_\text{kick} \approx v_\text{rec}$.
	This kick ensures a negligible overlap with the second excited band (smaller than 4\%).
	Although, in an ideal Hall system, all bulk excitations evolve periodically at the cyclotron  frequency $\omega_{\text{c}} = qB/M$ due to the harmonic spacing of successive Landau levels, this is not exactly the case in our system.
	We test this behavior by varying the strength of the excitation which relates to the magnitude of the velocity kick.
	As shown in \figref{fig_orbits_radii}d, we find that the trajectories cease to be closed and start to drift along the kick direction as the excitation strength exceeds $1.5\,\vr$  (see \figref{fig_orbits_radii}f).
	This regime corresponds to the onset of significant population of higher energy bands $n\geq 2$, which illustrates the non-harmonic spectrum of our system.

	It is important to note that the excitation protocol described so far is inefficient for large values of $p$, where the energy gap is much larger.
	In that regime, a quench of the coupling amplitude $\Omega$ leads to a more efficient overlap  with higher energy bands.
	This is shown in \figref{fig_orbits_radii}e, for the case of a sudden branching of the coupling strength to $\hbar\Omega = \Er$.
	The system initially at $p=-J\pr$ is then effectively coupled to higher energy bands and the bouncing on the hard wall characteristic of classical skipping orbits is clearly visible.


 \begin{figure*}[ht!]
\includegraphics[
draft=false,scale=1.,
trim={1mm 2mm 0 0.cm},
]{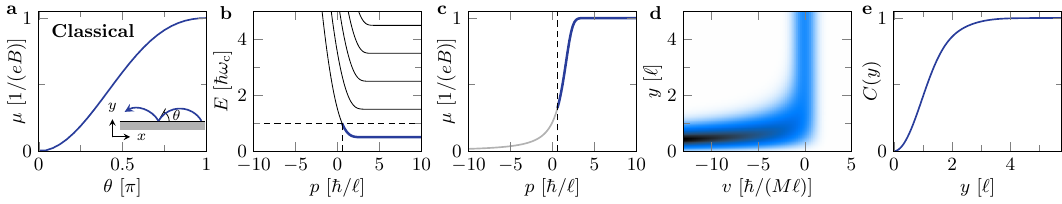}
\caption{
\textbf{Hall system in real dimensions.}
\textbf{a.} Variation of the Hall mobility for classical skipping orbits, depending on the angle of rebound on a hard wall.
The case of closed cyclotron orbits corresponds to $\theta=\pi$.
\textbf{b.} Dispersion relation of a quantum Hall system in a semi-infinite geometry $y>0$.
The blue line indicates the energy  branch used for the Chern marker calculation, defined by $E_0(p)<\hbar\omega_c$.
\textbf{c.} Hall mobility $\mu$ as a function of momentum $p$.
\textbf{d.} Local density of state in the $(v,y)$ plane.
\textbf{e.} Local Chern marker $C(y)$ for the energy branch defined in \textbf{b}.
\label{fig_hall_ribbon}}
\end{figure*} 

\smallskip
\noindent\textbf{Transverse drift in a Hall system}\\
	\noindent Our system is analogous to a Hall system in a ribbon geometry (see Supplementary information for a discussion in the case of a disk geometry).
	To understand the role of a sharp edge on the physical quantities measured in the main text, we consider an electronic Hall system in a semi-infinite geometry, described by the Landau Hamiltonian \eqref{eq_H_Landau}, written as
	\[
	\hat{H}=\frac{\hat{p}_y^2}{2M}+\frac{1}{2}M\omc^2(\hat{y}-\hat{p}_x\ell^2/\hbar)^2,
	\]
	with a hard-wall restricting motion to the half-plane  $y>0$.
	Here, we introduce the cyclotron frequency $\omc=eB/M$ and the magnetic length $\ell=\sqrt{\hbar/eB}$, assuming a magnetic field $B$ along $z$.

	We first consider semi-classical trajectories in the absence of external forces, which are either closed cyclotron orbits or skipping orbits bouncing on the edge, parametrized by the rebound angle $\theta$ (see \figref{fig_hall_ribbon}a).
	Applying a perturbative force $F$ along $y$ leads to a drift of cyclotron orbits of velocity $v_{\text{d}}=-F/eB$ along $x$, corresponding to a Hall mobility $\mu=1/eB$.
	For skipping orbits, the Hall drift can be expressed analytically as
	\begin{equation}\label{eq_H_Hall}
	\mu=\frac{1}{eB}\left[1-\left(\frac{\sin\theta}{\theta}\right)^2\right].
	\end{equation}
	The factor of reduction compared to cyclotron orbits, plotted in \figref{fig_hall_ribbon}a, smoothly interpolates between 1 for almost closed orbits ($\theta\rightarrow\pi$) and 0 for almost straight orbits ($\theta\rightarrow 0$).
	This behavior provides a simple explanation of the reduced Hall mobility of edge modes (see Fig.\,\ref{fig_hall}b).

	We extend this reasoning to the quantum dynamics in the lowest energy band.
	In a semi-infinite geometry, the eigenstates of the Hamiltonian (\ref{eq_H_Hall}) can be indexed by the momentum $p$ along $x$, and are expressed as \cite{de_bievre_propagating_2002}
	\begin{align}
	\psi_p(x,y)&=\frac{\E^{\I px/\hbar}}{\sqrt{2\pi\hbar}}\phi_p(y),\label{eq_psi_p}\\
	\phi_p(y)&\propto D_{\epsilon(p)-1/2}\left[\sqrt{2}(y-p\ell^2/\hbar)/\ell\right],\nonumber
	\end{align}
	where $D_\nu(z)$ is the parabolic cylinder function and $\epsilon(p)=E_0(p)/\hbar\omega_c$ is the reduced energy determined by the boundary condition $D_{\epsilon(p)-1/2}(-\sqrt{2}p\ell/\hbar)=0$ (see \figref{fig_hall_ribbon}b).
	By summing over all momentum states of the ground band, we compute the local density of state in $(v_x,y)$ coordinates plotted in  \figref{fig_hall_ribbon}d.
	Far from the edge $y\gg \ell$, the velocity distribution is a Gaussian centered on $v_x=0$, of rms width $\delta v_x=\hbar/(M \ell)$.
	The distribution is shifted to negative velocities when approaching the edge $y=0$, as expected for chiral edge modes.

	We now consider the Hall response of the system by studying the perturbative action of a force $F$ along $y$, described by the Hamiltonian
	\begin{align*}
	\hat{H}_p'&=\frac{\hat{p}_y^2}{2M}+\frac{1}{2}M\omc^2(\hat{y}-p\ell^2/\hbar)^2-F \hat{y},\\
	&=\frac{\hat{p}_y^2}{2M}+\frac{1}{2}M\omc^2\left(\hat{y}-\frac{p\ell^2}{\hbar}-\frac{F}{M\omc^2}\right)^2+\mathcal{E}(p),\\
	\mathcal{E}(p)&=-\frac{p F}{M\omc}-\frac{F^2}{2 M\omc^2}.
	\end{align*}
	We identify the perturbed Hamiltonian $\hat{H}_p'$ as $\hat{H}_{p+F/\omega_c}$, with an additional energy shift $\mathcal{E}(p)$.
	Assuming the system to remain in the ground band, the group velocity of a localized wavepacket becomes
	\[
	\langle \hat v \rangle' = v_0(p+F/\omc)-\frac{F}{M\omc},\quad v_0(p)=\frac{\dd E_0(p)}{\dd p}.
	\]
	Assuming a small force, we expand the velocity as $\langle \hat v \rangle'=v_0(p)-\mu(p) F$, with the mobility
	\begin{align*}
	\mu(p)&=\frac{1}{eB}\left(1-M\frac{\dd v_0}{\dd p}\right).
	\end{align*}
	This formula is analogous to the expression for the Hall mobility in our synthetic system.
	As shown in \figref{fig_hall_ribbon}c, it is close to the classical Hall drift in an infinite plane in the bulk mode region $p\gtrsim\hbar/\ell$, while it decreases towards zero in the edge mode region $p<0$.

	The overall response of an energy branch in the ground band can be obtained by summing the drifts of all populated eigenstates, such that the center of mass drift reads
	\[
	\langle v (t)\rangle' =\langle v (t)\rangle_0 - F\int\dd p\, n(p)\mu(p),
	\]
	where we assume the normalization $\int\dd p\, n(p)=1$ for the occupation number $n(p)$.
	We consider a uniform occupation of the lowest energy band, restricted to the energy branch $E_0(p)<\hbar\omc$, i.e. in the middle of the bulk gap to the first excited band in the bulk.
	This condition corresponds to momentum states $p>p^*\simeq0.54\hbar/\ell$ of the ground band.
	Assuming an upper momentum cutoff $p'$ in the bulk region, we obtain the Hall drift
	\[
	\langle v (t) \rangle'=\langle v (t)\rangle_0 -\frac{F}{eB}\left(1-M\frac{v_0(p')-v_0(p^*)}{p'-p^*}\right).
	\]
	As long as $p'\gg \hbar/\ell$, the second term can be neglected, and one recovers the Hall drift of a topological band of Chern number $\mathcal{C}=1$.

	We finally consider the local Hall response in the ground band, quantified by the local Chern marker \cite{bianco_mapping_2011}
	\[
	C(x,y)=2\pi\,\text{Im}\bra{x,y}[\hat{P}\hat{x}\hat{P},\hat{P}\hat{y}\hat{P}]\ket{x,y},
	\]
	where $\hat{P}$ projects on the considered branch of states and $\ket{x,y}$ are localized in $(x,y)$.
	The calculation of the Chern marker starts by decomposing position states into momentum states, as
	\[
	C(x,y)=\frac{2}{\hbar}\,\text{Im}\left[\int\dd p\,\dd q\,\E^{\I (p-q)x/\hbar}\phi_p(y)\phi_q^*(y)\tilde c(p,q)\right],
	\]
	where $\tilde c(p,q)\equiv\bra{\psi_p} \hat{x}\hat{P}\hat{y} \ket{\psi_q}$, which can be evaluated using the explicit form (\ref{eq_psi_p}) for momentum states as
	\[
	\tilde c(p,q)=\I \hbar \,\langle y\rangle_q \langle \phi_p|\phi_q\rangle\delta'(p-q),
	\]
	where $\langle y\rangle_q$ is the mean $y$ position in the wavefunction $\phi_q$.
	Using the general formula
	\begin{align*}
	&\int\dd u\,\dd v\,f(u,v)\delta'(u-v)\\&=\frac{1}{2}\int\dd u\,\dd v\,[\partial_vf(u,v)-\partial_uf(u,v)]\delta(u-v),
	\end{align*}
	we obtain the expression for the Chern marker
	\[
	C(x,y)=\int\dd p\,|\phi_p(y)|^2\frac{\dd \langle y\rangle_p}{\dd p}.
	\]
	The relation $p=M v_0+qB \langle y\rangle_p$ then leads to 
	\begin{equation}
	C(x,y)=\int\dd p\,|\phi_p(y)|^2\mu(p),
	\label{lcm}
	\end{equation}
	a relation analogous to the local Chern marker expression for our synthetic Hall system.
	We show in \figref{fig_hall_ribbon}e the Chern marker calculated for an energy branch $E(p)<\hbar\omega$, which is close to 1 for $y\gtrsim \ell$, and decreases towards zero when approaching the edge $y=0$, similarly to the decrease of the Chern marker close to the edges shown in Fig.~\ref{fig_hall}c.


\begin{figure*}[ht!]
\includegraphics[
draft=false,scale=1.,
trim={1mm 2mm 0 0.cm},
]{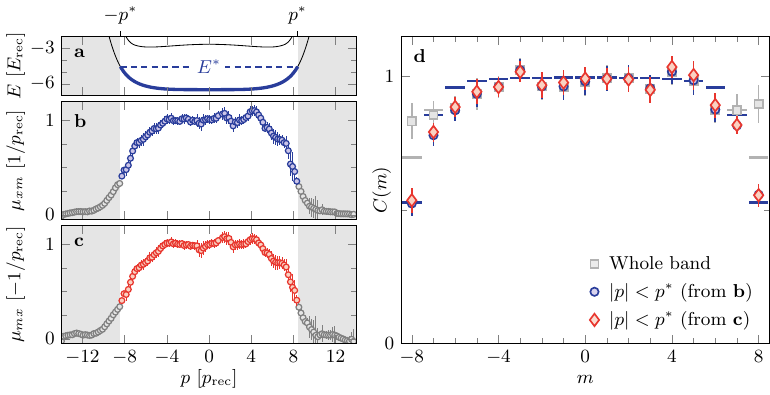}
\caption{
\textbf{Hall mobility and local Chern markers.}
\textbf{a.} Predicted dispersion relation for $\hbar\Omega = E_{\text{rec}}$.
The branch pictured in blue, chosen as $E(p) < E^{*}$ with $E^*$ at half the gap, is used for the computation of the local Chern marker.
\textbf{b.} Measured mobility in $x$ resulting from the application of a force along $m$, as presented in the main text.
The points in blue, corresponding to $|p| < p^*$ (white area), are the ones considered for the Chern marker presented in the main text (see Fig.\,\ref{fig_hall}).
\textbf{c.} Measured mobility in $m$ resulting from the application of a force along $x$.
As for \textbf{b}, the points in red are associated to momentum states lying below $E^*$.
\textbf{d.} Chern marker obtained from the measured mobility, using the whole energy branch ($-\infty < p < \infty$, gray squares, using data in \textbf{b}), or using the branch defined in \textbf{a} ($-p^* < p < p^*$).
For the latter, the blue dots correspond to the data in \textbf{b}, and are identical to Fig.\,\ref{fig_hall}.
The red diamonds correspond to the data in \textbf{c}.
Solid lines are theoretical values.
The error bars are the 1-$\sigma$ statistical uncertainty calculated from a bootstrap sampling analysis over typically 100 pictures (\textbf{b},\textbf{c}) and 1000 pictures (\textbf{d}).
\label{lcm_supp}
}
\end{figure*}

\smallskip
\noindent\textbf{Local Chern marker in synthetic dimension} \\
\noindent In the synthetic Hall system, the expression of the local Chern marker reads
	\begin{equation}\label{eq_Chern_marker}
	C(x, m) = 2\pi\,\text{Im}\bra{x,m}[\hat{P}\hat{x}\hat{P},\hat{P}\hat{J}_z\hat{P}]\ket{x,m}.
	\end{equation}
	Translation invariance along $x$ ensures that the Chern marker only depends on the coordinate $m$.
	In the main text, the notation $\ket{m}$ refers to an arbitrary $\ket{x,m}$ state, the choice of $x$ being irrelevant.
	The derivation of the Chern marker 
	\[
	C(m)=\int\text{d}p\,\Pi_m(p)\mu(p)
	\]
	is obtained following the same procedure as for a standard Hall system, discussed above.
	So far, we have only considered one component $\mu_{xm}$ of the mobility tensor -- the one that measures the drift along $x$ resulting from a force along $m$.
	One can also consider the other component, which quantifies the magnetization drift $\dd\langle \hat{J}_z\rangle/\dd t$ which results from a force $F_x$ along $x$.
	In linear response, it is defined as $\langle \hat{J}_z\rangle = \langle \hat{J}_z\rangle_0 - \mu_{mx}(p)F_xt$, where $\mu_{mx}$ explicitly designates the mobility component considered here, and $\langle \hat{J}_z\rangle_0$ is the unperturbed magnetization.
	Its expression is given by
	\begin{equation*}
		\mu_{mx} = -\frac{\dd}{\dd F_x}\frac{\dd \langle \hat{J}_z\rangle}{\dd t} = -\frac{\dd \langle \hat{J}_z(p)\rangle}{\dd p},
	\end{equation*}
	where we used $F_x = \dot{p}$ and the fact that $\langle \hat{J}_z\rangle_0$ is time-independent.
	The expression $\hat p=M\hat v+\pr \hat J_z$ allows to recover the relation $\mu_{mx} = -\mu_{xm}$ between the two transverse mobilities.

	We show in \figref{lcm_supp}b,c the measurements of both mobilities as a function of $p$, and find good agreement between them.
	We also present in
	\figref{lcm_supp}d the local Chern markers computed using the data of each mobility.

	In the main text, the Chern marker is evaluated over a branch of the ground band, below an energy threshold shown in \figref{lcm_supp}a (at half the cyclotron gap at $p = 0$).
	We also show the Chern marker computed using all momentum states (gray points).
	Compared to the restricted branch, we only find a discrepancy on the edges of the ribbon.
	In the region $-5 \leq m \leq 5$, the values are nearly identical, showing that the bulk topological response is insensitive to the momentum cutoff.

\begin{figure}[ht!]
\includegraphics[
draft=false,scale=1,
trim={2mm 2mm 0 0.cm},
]{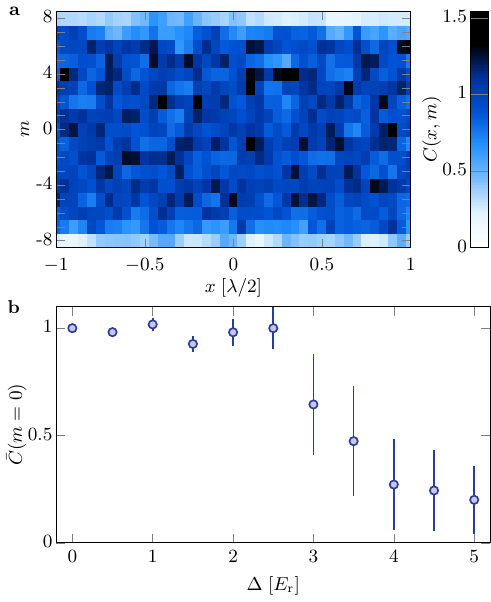}
\caption{
\textbf{Effect of disorder.}
\textbf{a.} Example of Chern marker distribution in the presence of disorder of strength $\Delta=\Er$.
\textbf{b.} Chern marker $\bar C(m=0)$ averaged over the region $|x|<\lambda/4$ as a function of the disorder strength $\Delta$.
Each point is the average of 100 disorder realizations, the error bar showing the standard deviation of the mean.
}
\label{fig_Chern_disorder}
\end{figure}

	We also evaluate theoretically the effect of disorder on the Chern marker.
	For this, we consider a finite-size system of length $L=5\lambda/2$, with periodic boundary conditions along $x$, and discretized on a grid $(x_n=n\delta x,m)$ of spacing $\delta x=\lambda/40$.
	The atom dynamics is described by the Hamiltonian \eqref{eq_H} with an additional disorder potential, taken as a random energy at each site $(x_n,m)$ drawn according to a normal distribution of rms $\Delta$.
	We calculate the energy spectrum and the local Chern marker using the definition (\ref{eq_Chern_marker}), where $\hat P$ projects on the eigenstates of energy $E<E^*$, with $E^*=\hbar\omc$ is the middle of the bulk gap.
	We show in \figref{fig_Chern_disorder}a an example of Chern marker distribution in the region $|x_n|<\lambda/2$ for a disorder strength $\Delta=\Er$.
	We define a coarse-grained average at the center of the synthetic dimension as
	\[
	\bar C(m=0)=\langle C(x_n,m=0)\rangle_{|x_n|<\lambda/4}.
	\]
	We show in \figref{fig_Chern_disorder}b the variation of $\bar C(m=0)$ with the disorder strength $\Delta$, averaged over 100 disorder realizations for each value of $\Delta$.
	We find that the central Chern marker is almost unchanged for disorder strengths $\Delta\lesssim 2\,\Er$, demonstrating the robustness of the Chern marker in the bulk of the sample.
	
\smallskip
\noindent\textbf{Abrikosov vortex lattices}\\
	\noindent The role of interactions in the ground band is assumed to be governed by a single parameter $g$ which describes contact interactions in both the real and the synthetic dimensions (see Supplementary information).
	We consider a gas of bosonic atoms with high filling fractions, for which the many-body ground state is well captured by mean-field theory.
	The system is described by  a spinor classical field $(\psi_m(x))$ (with $-J\leq m\leq J$), whose dynamics is governed by the Gross-Pitaevskii equation
	\begin{widetext}
	\[
	\I\hbar \dot\psi_m=\frac{\hbar^2}{2M}(\I\partial_x+K m)^2\psi_m
	-\hbar\Omega\!\left(\!\frac{\sqrt{J(J\!+\!1)\!-\!m(m\!+\!1)}}{2}\psi_{m-1}+\frac{\sqrt{J(J\!+\!1)\!-\!m(m\!-\!1)}}{2}\psi_{m+1}+\frac{m^2}{2J\!+\!3}\psi_m\!\right)+g|\psi_m|^2\psi_m.
	\]
	\end{widetext}

	From the phenomenology of Abrikosov vortex lattices, we expect the ground state to break translational invariance along the real dimension, with an unknown periodicity $L_0$.
	To find the period $L_0$, we numerically calculate  the ground state on a cylinder of circumference $L$, corresponding to periodic boundary conditions along the real dimension, by evolving the Gross-Pitaevskii equation in imaginary time.
	We find that the ground-state energy is minimized for a set of circumferences $L$, integer multiples of a length that we identify as $L_0$.

\begin{figure*}[ht!]
\includegraphics[
draft=false,scale=1,
trim={2mm 2mm 0 0.cm},
]{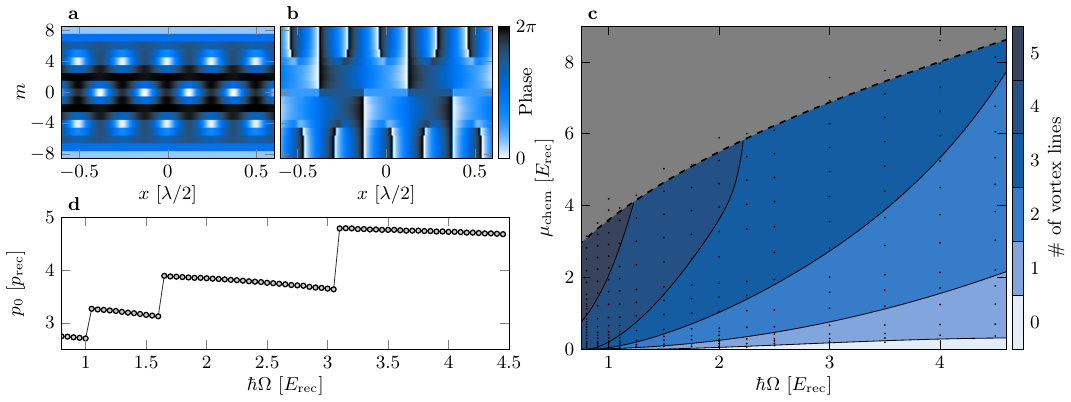}
\caption{
\textbf{Abrikosov vortex lattices.}
\textbf{a.} Ground state density profile and \textbf{b.} associated phase, for $\hbar\Omega = 3\Er$ and for $\mu_{\text{chem}}\approx 4\Er$.
The local minima of the density exhibit a phase winding around them, and thus correspond to quantum vortices.
\textbf{c.} Number of vortex lines as a function of the Raman coupling $\Omega$ and the chemical potential $\mu_{\text{chem}}$.
The dots identify the configurations for which a simulation was realized.
The color encodes the number of vortex lines that characterizes the low-energy vortex lattice configuration.
The phase separation lines are guides to the eye.
The dashed line identifies the gap to the first excited band above which the atoms significantly occupy higher Landau levels.
\textbf{d.} Momentum $p_0$ associated to the spontaneous breaking of the translational invariance resulting from the appearance of a vortex lattice, as a function of $\Omega$.
The points were taken at a chemical potential corresponding to half the gap.
}
\label{abrikosov_vortex_lattice}
\end{figure*}

	The thermodynamic properties are determined by the coupling $\Omega$ and the interaction energy scale $g\langle n\rangle$, where $\langle n\rangle$ is the mean atom density, or equivalently by the chemical potential $\mu_{\text{chem}}$.
	Here we explore situations in which the chemical potential lies in the gap between the LLL and the first excited band (see \figref{abrikosov_vortex_lattice}b.).

	For large enough interactions, we always find ground state configurations in the shape of Abrikosov triangular vortex lattices, such as the ones presented in the main text (see Fig.\,\ref{fig_vortex}).
	We give in \figref{abrikosov_vortex_lattice}a,b another example of such ground state, represented here by both the density profile and the phase associated to the wavefunction.
	Around each local minimum of the density, the phase profile is reminiscent of the phase winding of a quantum vortex in a continuous 2D system.

	The hard walls in the synthetic dimension have a strong impact on the vortex lattice geometry.
	We distinguish the different configurations by counting the number of vortex lines along $x$.
	For example, in \figref{abrikosov_vortex_lattice}a we identify a configuration made of $3$ vortex lines.
	The phase diagram, shown in \figref{abrikosov_vortex_lattice}c, shows a large variety of vortex configurations.
	Typically, the distance between lines is set by the magnetic length $\ell_m$ in the synthetic dimension.
	The reduction of the number of vortex lines with $\Omega$ is thus explained by the increase of $\ell_m$.
	Similarly to type-II superconductors in a confined geometry, the different vortex configurations are separated by first-order transition lines.

	The observation of such vortex lattices demands a high-resolution \emph{in situ} imaging resolved in $m$ space.
	However, the spontaneous breaking of translational symmetry can be revealed in the momentum distribution -- accessible in standard time-of-flight experiments -- via the occurrence of Bragg diffraction peaks at multiples of the momentum $p_0=h/L_0$.
	As shown in \figref{abrikosov_vortex_lattice}d, the expected variation of $p_0$ with the coupling $\Omega$ indirectly reveals the occurrence of phase transitions between different vortex configurations.


\smallskip
\noindent\textbf{Data availability}\\
\noindent 
Source data, as well as other datasets generated and analyzed during the current study are available from the corresponding author upon request.

\smallskip
\noindent\textbf{Code availability}\\
\noindent 
The source code for the numerical simulations of the Abrikosov vortex lattices and the Lauglin states are available from the corresponding author upon request.

\phantom{a}\\
\phantom{a}\\
\phantom{a}\\
\phantom{a}\\
\phantom{a}\\
\phantom{a}\\
\phantom{a}\\
\phantom{a}\\
\phantom{a}\\
\phantom{a}\\
\phantom{a}\\
\phantom{a}\\
\phantom{a}\\
\phantom{a}\\
\phantom{a}\\
\phantom{a}\\
\phantom{a}\\
\phantom{a}\\
\phantom{a}\\
\phantom{a}\\
\phantom{a}\\
\phantom{a}\\
\phantom{a}\\
\phantom{a}\\
\phantom{a}\\
\phantom{a}\\
\phantom{a}\\
\phantom{a}\\
\phantom{a}\\
\phantom{a}\\
\phantom{a}\\
\phantom{a}\\
\phantom{a}\\
\phantom{a}\\
\phantom{a}\\
\phantom{a}\\
\phantom{a}\\
\phantom{a}\\
\phantom{a}\\
\phantom{a}\\
\phantom{a}\\
\phantom{a}\\
\phantom{a}\\
\phantom{a}\\
\phantom{a}\\
\phantom{a}\\
\phantom{a}\\
\phantom{a}\\
\phantom{a}\\
\phantom{a}\\
\phantom{a}\\

\newpage

\section*{Supplementary information}

\section{Atom lifetime in the ground band}
\begin{figure}[t!]
	\includegraphics[
	draft=false,scale=1,
	trim={4mm 2mm 0 0.cm},
	]{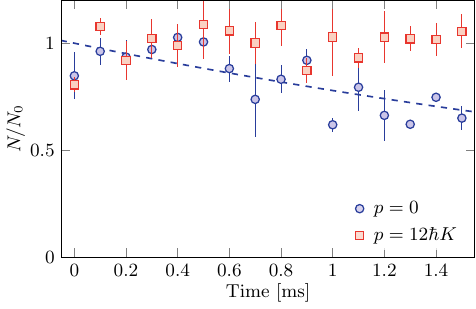}
	\caption{
	\textbf{Lifetime in the ground band.}
	Remaining atom number $N$ for a state prepared at $p = 0$ (blue) and $p = 12\hbar K$ (red), in the presence of the Raman coupling at $\hbar\Omega = \Er$.
	The atom numbers are normalized by $N_0$, the value at the origin of the fit (dashed line) which indicates a $1/\text{e}$ time constant of $\SI{4}{\ms}$.
	The error bars are the 1-$\sigma$ standard deviation of typically 5 measurement repetitions.
	\label{fig_lifetime}}
	\end{figure}

	\noindent As explained in the main text (Methods), we experimentally find that we can adiabatically prepare the ground state for any value of $p$, which suggests the absence of significant inelastic processes leading to higher band population.
	We confirmed such an observation by measuring the atom lifetime, following a preparation in the states $p = 0$ and $p = 12\hbar K$.
	Experimentally, this amounts to measuring the remaining number of atoms after a holding time, in the presence of the Raman coupling at $\hbar\Omega = \Er$ (see \figref{fig_lifetime}).

	We measure a lifetime of approximately \SI{4}{ms}, for $p=0$, which cannot be simply attributed to incoherent Rayleigh scattering processes associated to the Raman coupling.
	Indeed, we estimate numerically the spontaneous emission rate to be, at most, on the order of \SI{15}{\s^{-1}}, corresponding to a timescale larger than $\sim\SI{60}{\ms}$, an order of magnitude larger than the one reported in \figref{fig_lifetime}.
	However, we notice that the lifetime is significantly larger for $p=12 \hbar K$, which suggests the presence of dipolar relaxation \cite{burdick_fermionic_2015}.
	We find numerically that the typical loss rate associated to the dipolar relaxation, for a density of $ \SI{e13}{\cm^{-3}}$, varies between $\sim\SI{70}{\s^{-1}}$ and $\sim\SI{230}{\s^{-1}}$ (depending on the spin state), which is consistent with our measurements.
	This loss mechanism could be inhibited in a modified protocol, in which the Hamiltonian \eqref{eq_H} (main text) is realized in the absence of external magnetic field, which requires a different laser configuration.

\section{Emergence of Landau levels}

	\begin{figure*}[ht!]
	\includegraphics[
	draft=false,scale=1,
	trim={4mm 2mm 0 0.cm},
	]{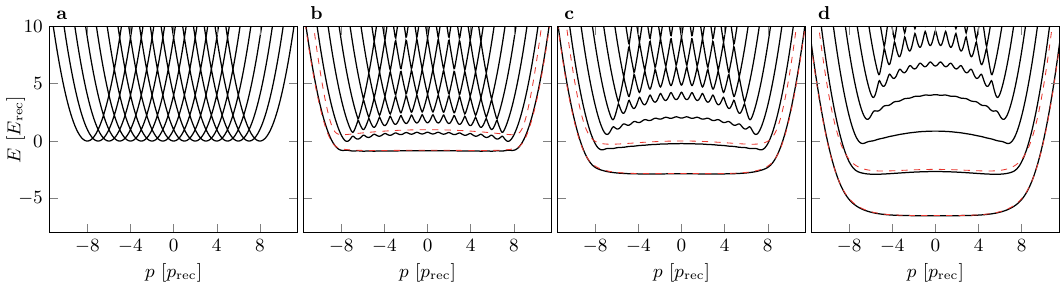}
	\caption{
	\textbf{Emergence of synthetic Landau levels.}
	\textbf{a-d.} Dispersion relations for coupling strengths $\hbar\Omega/\Er = 0, 0.2, 0.5$ and $1$ (solid lines).
	The dashed red lines correspond to the first two energy bands of a semi-classical theory, including first-order quantum corrections.
	\label{fig_dispersions}}
	\end{figure*}
	
	\noindent In \figref{fig_dispersions}, we show the dispersion relation of the Hamiltonian \eqref{eq_H} (see main text) calculated for different couplings $\Omega$.
	In the absence of the light coupling, $\Omega=0$, the Hamiltonian reduces to the kinetic energy term $(\hat p-\pr \hat J_z)^2/(2M)$, leading to $2J+1$ parabolas shifted along $p$.
	All energy crossings become avoided for $\Omega\neq0$, leading to flattened energy bands akin to Landau levels.
	Achieving a flat ground band dispersion in the bulk region requires couplings that are large enough ($\hbar\Omega\gtrsim0.2\,\Er$) to reduce short-$p$ oscillations, while still being sufficiently small ($\hbar\Omega\lesssim\Er$) to minimize longer-scale curvature.

	The large spin $J=8$ allows for a simplified semi-classical description, where the spin is represented by a point on the generalized Bloch sphere, parametrized by its spherical angles $(\theta,\phi)$.
	The  spin projection  is mapped on a continuous variable $m=J\cos\theta$, with the azimuthal angle $\phi$ being its conjugated variable (up to a factor $\hbar$).
	The semi-classical Hamiltonian corresponding to the quantum Hamiltonian \eqref{eq_H} (see main text) then reads
	\[
	\mathcal{H}_p=\frac{(p-\hbar K  m)^2}{2M}-\hbar\Omega\left(\sqrt{J^2-m^2}\cos\phi+\frac{m^2}{2J+3}\right).
	\]
	This energy functional being minimized for $\phi=0$, one can assume $\phi\ll 1$ and obtain a low-energy expansion 
	\begin{align*}
	\mathcal{H}_p&=\frac{(p-\hbar K  m)^2}{2M}+\frac{\hbar^2\phi^2}{2M'(m)}+V(m),\\
	M'(m)&=\hbar/\left(\Omega\sqrt{J^2-m^2}\right),\\
	V(m)&=-\hbar\Omega\left(\sqrt{J^2-m^2}+\frac{m^2}{2J+3}\right),
	\end{align*}
	which is exactly the Landau Hamiltonian \eqref{eq_H_Landau} (see main text), albeit with a position-dependent mass $M'(m)$ and a confining potential $V(m)$ in the synthetic dimension.
	The divergence of the mass $M'(m)$ for $|m|\rightarrow J$ leads to an effective hard-wall condition.

	In the middle of the bulk, at $m=0$, in the semi-classical model, we deduce the cyclotron frequency $\omega_c=\sqrt{2J \Omega\Er/\hbar}$, and we expect a value $\omega_c=4\Er/\hbar$ for $\hbar\Omega=\Er$, which is close to the exact value $\omega_c=3.84\Er/\hbar$ (see inset of \figref{fig_orbits}, see main text).
	We also infer the expressions for magnetic lengths 
	\[
	\ell_m=\sqrt[4]{\frac{J\hbar\Omega}{2\Er}}\quad\text{and}\quad \ell_x=\frac{1}{K \ell_m}
	\]
	in the synthetic and real dimensions respectively.
	These lengths are the characteristic sizes of the quantum vortices shown in Fig.~\ref{fig_vortex}b (see main text).
	For the coupling $\hbar\Omega=\Er$ used in the simulations, we obtain magnetic lengths $\ell_m\simeq1.41$ and $\ell_x\simeq 0.11\lambda/2$.

	The approximate analogy between the Hamiltonians \eqref{eq_H} and \eqref{eq_H_Landau} (see main text) can also be inferred using quantum operators, as we explain now assuming $p\simeq 0$ for simplicity.
	In that case, we expect the system to be polarized in $m=J$ along $x$, such that the commutator
	\begin{align*}
			[ \hat J_z, \hat J_y] &= -\I \hat J_x \simeq -\I J
	\end{align*}
	is a $c$-number.
	The operator $-\hat J_y/J$ is then canonically conjugated to the spin projection $\hat J_z$.
	We then use the Holstein-Primakoff approximation at second order to express the spin projection $\hat J_x$ as
	\[
	\hat J_x=J+\frac{1}{2}-\frac{\hat J_y^2+\hat J_z^2}{2J},
	\]
	leading to the Hamiltonian
	\[
	\hat H=\frac{(p-\pr\hat J_z)^2}{2M}+\frac{\hbar\Omega}{2J}\hat J_y^2+\frac{3 \hbar\Omega}{2J(2J+3)}\hat J_z^2-\hbar\Omega\left(J+\frac{1}{2}\right).
	\]
	This Hamiltonian corresponds to the Landau Hamiltonian \eqref{eq_H_Landau} (see main text), with an additional $\hat J_z^2$ term.
	This approximation can be generalized to all values of momentum $p$.
	We show in \figref{fig_dispersions} the first two energy bands calculated within this approximation, together with the exact spectrum.

\section{Ground band flatness and symmetry}
\begin{figure}[t!]
	\includegraphics[
	draft=false,scale=0.95,
	trim={2mm 2mm 0 0.cm},
	]{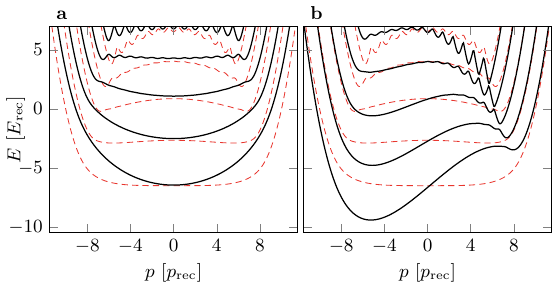}
	\caption{
	\textbf{Ground band flatness.}
	\textbf{a.} Dispersion relation after addition of a harmonic confining potential proportional to $J_z^2$ (solid lines),  which cancels $V(\hat J_z)$ of Eq.~\eqref{eq_H} (see main text).
	\textbf{b.} Dispersion relation in case of using Raman laser beams in a $\sigma\operatorname{-}\pi$ configuration (solid lines).
	In both panels, the red dashed lines correspond to the dispersion relation of the Hamiltonian \eqref{eq_H} (see main text).
	\label{fig_flatness}}
	\end{figure}

	\noindent The flatness of the ground band and the uniformity of the cyclotron frequency in our system arise from the partial cancellation of the dispersive effects from the kinetic term $-\hbar\Omega \hat J_x$ and the light shift term $V(\hat J_z)=-\hbar\Omega\hat J_z^2/(2J+3)$ in the Hamiltonian \eqref{eq_H} (see main text).
	To illustrate this, we show the effect of removing the term $V(\hat J_z)$ from the Hamiltonian in \figref{fig_flatness}a -- clearly resulting in a highly dispersive lowest band.
	This is equivalent to an additional harmonic confining potential along the synthetic dimension, which is achievable in practice using an additional laser beam linearly polarized along $z$, and far-detuned from both Raman beams.

	Furthermore, the symmetry of the ground band about $p=0$ depends on the polarization of the two coupling lasers.
	The Raman transition scheme we use (Fig.~\ref{fig_scheme}a, see main text) could also be achieved using a polarization $u_1 = y = (\hat{\sigma}_{+} - \hat{\sigma}_{-})/\sqrt{2}\I$ for laser 1 and $u_2 = z = \hat{\pi}$ for laser 2.
	However, this $\sigma\operatorname{-}\pi$ configuration results in a highly asymmetric dispersion relation, as shown in \figref{fig_flatness}b.
	In our experimental scheme, we recover symmetric bands by allowing equal contributions from $\sigma\operatorname{-}\pi$ and $\pi\operatorname{-}\sigma$ arrangements, which corresponds to having orthogonal linear polarizations at $\SI{45}{\degree}$ to the $z$-axis.
	Imperfections in the orientation of the polarizations could explain the slight asymmetries we measure in velocity and the Hall mobility.
	
\section{Chern marker in a Hall disk}
\begin{figure}[t!]
	\includegraphics[
	draft=false,scale=1,
	trim={2mm 2mm 0 0.cm},
	]{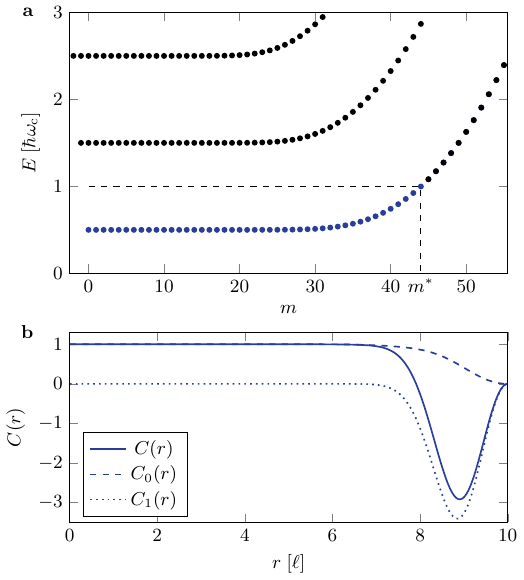}
	\caption{
	\textbf{Spectrum and Chern marker in a Hall disk.}
	\textbf{a.} Energy spectrum of a Hall disk of radius $R=10\,\ell$, indexed by the angular momentum projection $m$.
	The blue dots correspond to the states of energy $E_{n,m}<\hbar\omc$ considered for the Chern marker evaluation.
	\textbf{b.} Chern marker $C(r)$ in the same geometry, together with the two contributions $C_0(r)$ and $C_1(r)$.
	}
	\label{fig_disk}
	\end{figure}

	\noindent We extend the discussion of the main text (Methods) to the transverse response properties of a Hall system confined in a finite area, taking the example of a disk geometry.
	Writing the vector potential in the symmetric gauge, the Schr\"odinger equation reads
	\[
	\I\partial_t\psi=\frac{\hbar^2}{2M}\left[-\frac{1}{r}\partial_r(r\partial_r)+\left(\frac{-\I\partial_\theta}{r}- \frac{r}{2\ell^2}\right)^2\right]\psi
	\]
	in polar coordinates.
	Its eigenstate wavefunctions $\psi_{n,m}(r,\theta)=\phi_{n,m}(r)\E^{\I m\theta}$, indexed by an integer $n\in\mathbb{N}$ and the angular momentum projection $m\in\mathbb{Z}$, are solutions of the radial equation
	\[
	E_{n,m}\phi_{n,m}=\frac{\hbar^2}{2M}\left[-\frac{1}{r}\partial_r(r\partial_r)+\left(\frac{m}{r}- \frac{r}{2\ell^2}\right)^2\right]\phi_{n,m}.
	\]
	For $R\rightarrow\infty$, the states for a given $n$ are degenerate and form the $n^{\text{th}}$ Landau level.
	The wavefunction $\phi_{n,m}(r)$ takes significant values around the radius $r\simeq\sqrt{2m}\ell$.
	For a finite disk of radius $R$, we thus expect the states $\phi_{n,m}$ to remain almost degenerate for $m\lesssim(R/\ell)^2/2$.
	We show in \figref{fig_disk}a the energy spectrum for a disk of radius $R=10\,\ell$, consistent with this expectation.

	We now consider the transverse response of a system in which the states of energy $E_{n,m}<\hbar\omc$ (i.e. with $n=0$ and $m\leq m^*$) are uniformly occupied (see \figref{fig_disk}a).
	The local Chern marker can be expressed in terms of the radial wavefunctions as
	\begin{align*}
	C(\mathrm{r})&=C_{0}(r)+C_1(r),\\
	C_{0}(r)&=\sum_{m\leq m^*}\mu_m|\phi_{0,m}(r)|^2,\\
	C_{1}(r)&=\frac{|\phi_{0,m^*}(r)|^2+|\phi_{0,m^*+1}(r)|^2}{4}|\langle \phi_{0,m^*}|\hat r |\phi_{0,m^*+1}\rangle|^2,
	\end{align*}
	where we introduce
	\[
	\mu_m=\frac{1}{2}(|\langle \phi_{0,m}|\hat r |\phi_{0,m+1}\rangle|^2-|\langle \phi_{0,m}|\hat r |\phi_{0,m-1}\rangle|^2).
	\]
	The first term $C_0(r)$ is the sum of contributions from all occupied orbitals, analogously to the equation \eqref{lcm} (Methods) obtained in the half-plane geometry.
	It remains close to one in the bulk, and decreases to zero close to the edge $r=R$ over a length scale $\ell$.
	The term  $C_1(r)$ remains negligible in the bulk, and takes negative values of order $R/\ell\gg 1$ close to the edge.
	One checks that the spatial averages of the two terms are opposite, such that 
	\[
	\int\dd^2 r \,C(r)=0,
	\]
	as required for the Chern marker on a finite geometry \cite{bianco_mapping_2011}.
	Such a zero average does not occur in the experimental system, since the atom dynamics is not confined in the real dimension $x$.

\section{Interactions in the lowest energy band}

	\begin{figure*}[t!]
	\includegraphics[
	draft=false,scale=1,
	trim={2mm 2mm 0 0.cm},
	]{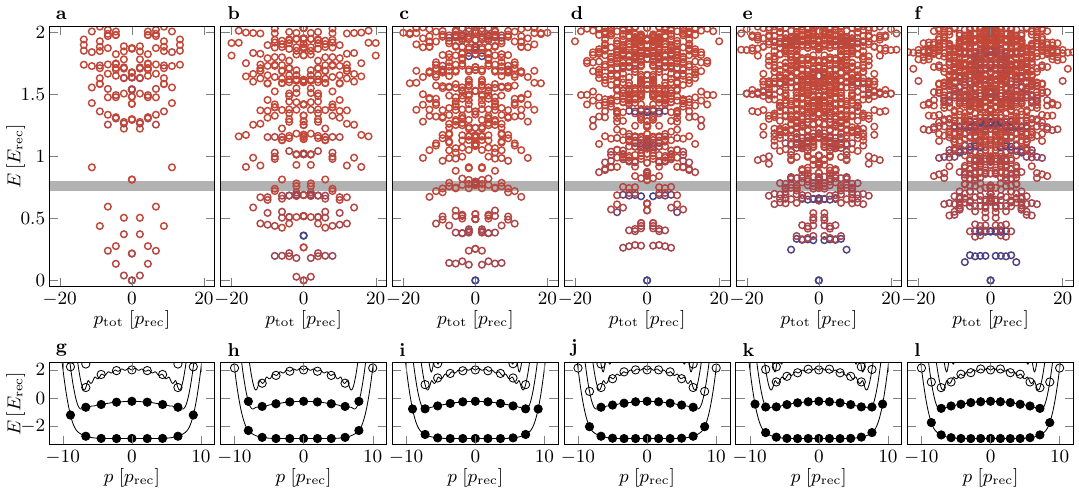}
	\caption{
	\textbf{Laughlin-like states at low filling.}
	\textbf{a}-\textbf{f}.
	Many-body energy spectra calculated for $N_{\text{at}}=5$ atoms on cylinders, of circumferences $L=0.45,0.5,0.55,0.6,0.65,0.7\,\lambda/2$, respectively.
	We identify a Laughlin-like state with very small interaction energy and at zero momentum, separated in energy from other eigenstates for $L\geq 0.55\,\lambda/2$.
	For $L=0.7\,\lambda/2$ we recognize an energy branch of edge excitations of the Laughlin state.
	The gray area marks the bulk gap of the Laughlin state $\Delta=0.60(3)\,U$ expected  in dispersionless Landau levels and in the thermodynamic limit \cite{regnault_quantum_2003}.
	The color scale is identical to the one used in Fig.~\ref{fig_vortex}c. \textbf{g}-\textbf{l}.
	Single-particle states included in the simulations are shown as solid dots.
	\label{fig_fqh}}
	\end{figure*}
	
	\noindent Interactions are typically short-ranged in atomic gases.
	In our system, interactions are thus local in $x$, but they can occur between any pair of spin projections $m_1$, $m_2$, corresponding to highly long-range interactions along the synthetic dimension.
	To recover short-range interactions, we propose to spatially separate the different $m$ states using a magnetic field gradient oriented along another direction $z$.
	The resulting system is thus constituted of $2J + 1 = 17$ one-dimensional tubes, offset in position along $z$.
	For a transverse confinement of frequency $\omega_z=2\pi\times \SI{1}{kHz}$, corresponding to a ground-state extent $\Delta z=\sqrt{\hbar/M\omega_z}\simeq\SI{250}{nm}$, the interactions become short-ranged for moderate magnetic field gradients $\nabla B\gtrsim \SI{30}{G/cm}$.
	For such gradients, the distance between subsequent tubes is $\gtrsim\SI{320}{\nm}$, such that the spatial overlap of neighbouring wavefunctions is negligible, thus making contact  interactions only possible between atoms within the same tube.
	At such spatial separations, the long-range dipole-dipole interactions (DDI) furthermore remains limited to a few tens of \si{Hz} ($h = 1$), which is also much weaker than the contact interaction within each tube.

	The interaction between atoms in a given $m$ state is then described by a short-range potential $g_m\delta(x_1-x_2)$ with coupling constant $g_m$, proportional to the $s$-wave scattering length $a_m$.
	At low magnetic field, rotational symmetry ensures that  $a_m=a_{-m}$, such that interactions are described by $J+1$ independent scattering lengths.
	While the $a_m$ constants are uniform between all nuclear spin levels for two-electron atoms, we do not expect such a SU($N$) symmetry for lanthanide atoms such as dysprosium, for which only the coupling constant $a_8=140(20)\,a_0$ has been measured \cite{tang_$s$-wave_2015}.
	All the other $a_m$ constants remain unknown and we plan to investigate them in the future.
	Nonetheless, if all values $a_m$ are positive, the system will be protected from collapse, making many-body phases experimentally accessible.

	In lanthanide atoms, interactions between magnetic dipoles enrich the situation discussed above.
	These interactions offer an additional degree of freedom that could be used to stabilize the system in case of attractive $s$-wave interaction channels.

	For simplicity, we neglect dipolar interactions in the numerical simulations presented in the main text (Methods) and consider  that all scattering lengths are equal and positive, such that the interaction potential reads $g\,\delta(x_1-x_2)\delta_{m_1,m_2}$.
	For such contact interactions, one expects interactions restricted to the lowest Landau level to reduce to a single Haldane pseudo-potential \cite{haldane_fractional_1983,cooper_rapidly_2008}
	\[
	U=\frac{g}{4\pi}\frac{1}{\ell_x\ell_m}=\frac{gK}{4\pi}.
	\]

\section{Laughlin-like state at low filling}
	
	\noindent We consider in this section bosonic atoms at low filling fractions, for which one expects strongly-correlated ground states.
	We calculate the many-body spectrum of this system using exact diagonalization.
	The stability of Laughlin-like quantum states is based on limited energy dispersion in the ground band, which is improved by considering a coupling $\hbar\Omega=0.5\,\Er$, i.e. half of the value used in the experiment.
	We use a cylindrical geometry  to avoid edge effects along $x$, and restrict the basis of single-particle levels to an energy $E=3\,\Er$ above the single-particle ground state, which includes bulk states of the ground and first excited bands, with a few edge modes depending on the circumference $L$ of the cylinder (see \figref{fig_fqh}g-l).
	We calculate the energy spectrum of $N_{\text{at}}$  bosonic atoms interacting at short range, with an interaction strength $U=\Er$.
	The many-body eigenstates are indexed by the total momentum $p_{\text{tot}}$, a conserved quantity that permits us to subdivide the Hilbert space into independent sectors, limiting the involved matrices to dimensions less than $4000$.

	We show in \figref{fig_fqh}a-f six energy spectra calculated for cylinder circumferences in the range $L=0.45-0.7\,\lambda/2$.
	For $L=0.45\,\lambda/2$, the ground state does not exhibit any of the characteristics of the Laughlin state: sizeable interaction energy $E_{\text{int}}\simeq0.03\,\Er$ and phonon-like low-energy excitations.
	This behavior stems from the limited number  of single-particle orbitals at low energy $N_{\textrm{orb}}=7$, smaller than the  number $N_{\textrm{orb}}=2N_{\text{at}}-1=9$ of distinct orbitals involved in the Laughlin wavefunction \cite{laughlin_anomalous_1983,rezayi_laughlin_1994}.
	For circumferences  $L\geq0.55\,\lambda/2$, we find a ground state with a very small interaction energy $E_{\text{int}}\simeq1.5\times10^{-4}\,\Er$, indicating anti-bunching between atoms, as expected for the Laughlin state.
	This state is separated from excited levels by an energy gap of maximum value $\simeq0.27\,\Er$ (reached for $L=0.63\,\lambda/2$), without featuring a low-energy phonon branch.
	For longer circumferences $L\geq\lambda/2$, the low-energy excitations also exhibit very small interaction energy, as expected for edge excitations of the Laughlin state occurring for a number of low-energy orbitals $N_{\textrm{orb}}>2N_{\text{at}}-1$ \cite{wen_theory_1992,soule_edge_2012}.
	
\newbibstartnumber{44}
\putbib

\end{bibunit}

\end{document}